\documentclass[prb,twocolumn,superscriptaddress]{revtex4}

\usepackage{amsmath} 
 
\usepackage{graphicx} 
\usepackage{times}

\begin{document}

 
\title{Topological Entanglement 
Entropy from the Holographic Partition Function}

\author{Paul Fendley} 
\affiliation{Department of Physics, University of Virginia, 
Charlottesville, VA 22904-4714} 
\author{Matthew P.A. Fisher} 
\affiliation{Kavli Institute for Theoretical Physics, 
University of California, Santa Barbara, CA 93106-4030} 
\author{Chetan Nayak} 
\affiliation{Microsoft Project Q, 
University of California, Santa Barbara, CA 93106-4030} 
\affiliation{Department of Physics and Astronomy, 
University of California, Los Angeles, CA 90095-1547}

\date{September 4, 2006}

\begin{abstract} 
We study the entropy of chiral 2+1-dimensional topological phases,
where there are both gapped bulk excitations and gapless edge
modes. We show how the entanglement entropy of both types of
excitations can be encoded in a single partition function. This
partition function is holographic because it can be expressed entirely
in terms of the conformal field theory describing the edge modes. We
give a general expression for the holographic partition function, and
discuss several examples in depth, including abelian and non-abelian
fractional quantum Hall states, and $p+ip$ superconductors. We extend
these results to include a point contact allowing tunneling between
two points on the edge, which causes thermodynamic entropy associated
with the point contact to be lost with decreasing temperature.  Such a
perturbation effectively breaks the system in two, and we can identify
the thermodynamic entropy loss with the loss of the edge entanglement
entropy.  {}From these results, we obtain a simple interpretation of
the non-integer `ground state degeneracy' which is obtained in
1$+$1-dimensional quantum impurity problems: its logarithm is a
2$+$1-dimensional topological entanglement entropy.
\end{abstract}

\maketitle 
 

\section{Introduction} 
 
Entanglement is one of the characteristic peculiar features of quantum 
mechanics. It lay at the heart of the debates between Bohr and 
Einstein, Podolski, and Rosen, and it is essential for the Bell's 
inequality violation which distinguishes quantum mechanics from 
classical hidden variables theories.  Entanglement between logical 
qubits is a resource for quantum computation; entanglement between 
physical qubits and the environment is a threat to quantum 
computation; entanglement between physical qubits is essential for 
error correction. 
 
Entanglement entropy is one measure of entanglement. 
If a system can be subdivided into two subsystems $A$ 
and $B$, then even if the whole system is in a pure quantum 
state $|\Psi\rangle$, subsystem $A$ will be in a mixed 
state with density matrix $\rho$ obtained by tracing 
out the subsystem $B$ degrees of freedom: 
\begin{equation} 
\rho^{}_{A} =\text{tr}_{B}\bigl( \left | \Psi \right \rangle \! 
\left\langle \Psi \right | \bigr) 
\end{equation} 
The entropy of this density matrix will be zero if 
the state $|\Psi\rangle$ is the direct product of a pure 
state for subsystem $A$ with a pure state for subsystem $B$ 
or, equivalently, if the matrix ${\rho^{}_A}$ has only a single 
non-zero eigenvalue. If the entropy 
\begin{equation} 
\label{eqn:entanglement-entropy-def} 
{{\cal S}_A} = - \text{tr}\!\left({\rho^{}_A}\ln {\rho^{}_A}\right) 
\end{equation} 
is non-zero, then the subsystems $A$ and $B$ are 
entangled. 
 
The entanglement entropy ${\cal S}_A$ is a measure of 
the correlations between the degrees of freedom 
of the $A$ and $B$ subsystems. In a $1$+$1$-dimensional 
quantum system, the dependence of ${\cal S}_A$ 
on the length $L$ of subsystem $A$ can be used 
to distinguish critical from non-critical systems: 
it remains finite as the length increases, except 
at criticality, where it diverges logarithmically with 
a universal coefficient \cite{Holzhey94,Calabrese04}. 
Of course, in this case, there are other measures of criticality, 
such as the power-law decay of correlation functions. 
In a gapped system, the leading 
term in the entanglement entropy is proportional 
to the surface area of the boundary. The coefficient 
is cutoff-dependent. However,  
in a $2+1$-dimensional system in 
a gapped topological phase, the first subleading term 
is universal and independent of the size or 
shape of $A$ \cite{Kitaev05,Levin05}: 
\begin{equation} 
\label{eqn:topo-entanglement} 
{{\cal S}_A} = aL - \ln {\cal D} + \ldots 
\end{equation} 
It is not immediately obvious that ${\cal D}$ in the above equation
(\ref{eqn:topo-entanglement}) is uniquely defined since $a$ is a
cutoff-dependent coefficient (in two spacetime dimensions, the
`surface area' is the length of the boundary).  However, by dividing a
system into three or more subsystems and forming an appropriate linear
combination of the resulting entanglement entropies, the length term
can be canceled, leaving only the universal term
\cite{Kitaev05,Levin05}. Such a construction can be used to give a
more precise definition of the topological entanglement entropy, but
its essential meaning is captured by (\ref{eqn:topo-entanglement}).
The quantity ${\cal D}$ is the {\it total quantum dimension} of the
topological phase, which we define in section
\ref{sec:quantum-dimensions}.
 
This is potentially a very useful probe of a 
topological phase. In such a phase, 
all correlation functions are topologically-invariant 
at distances longer than some finite correlation length 
and energies lower than a corresponding energy scale. 
Hence, in order to identify such a state, 
one must examine the ground state degeneracy 
on higher-genus surfaces or the braiding properties 
of quasiparticle excitations. The entanglement entropy 
gives us a handle on this topological structure
stemming from the ground state wavefunction alone.  
 
Topological quantum field theories (TQFTs) 
were widely studied in the string-theory community in 
the late '80s and early '90s, following 
Witten's celebrated work \cite{Witten89} on Chern-Simons 
field theory, conformal field theory (CFT), and the 
Jones polynomial of knot theory \cite{Jones85}. 
There has been a recent revival of interest 
in TQFTs due to the possible emergence of 
topological phases in electronic condensed-matter 
systems and the proposed use of 
such phases as a platform for fault-tolerant quantum 
computation \cite{Kitaev97,Physics-Today}. 
The above definition of a topological phase 
can be compactly (though tautologically) 
restated as a phase of matter for which the long-distance
effective field theory is a topological field theory. 
The entanglement entropy is a way of extracting 
one of the basic parameters of a topological field theory, 
its quantum dimension.


A chiral TQFT in $2$+$1$ dimensions is related to
$1$+$1$-dimensional rational conformal 
field theory (RCFT) in several ways. When the TQFT is defined 
on a manifold with boundary, the boundary degrees 
of freedom form an RCFT. The topological invariance of 
the theory means that there is no distinction between spacelike 
and timelike boundaries. Hence, the RCFT describes both 
(a) the ground state wavefunction(s) at some $(2$+$0)$-d 
equal-time slice when spacetime is $M\times I$, for 
a compact surface $M$ and time interval $I$, 
{\em and} (b) the dynamics of the $(1$+$1)$-d 
boundary when spacetime is $X\times R$, where 
$X$ is a surface with boundary. 
In the RCFT, the topological braiding/fusion 
structure of the TQFT is promoted to a holomorphic 
structure with branch cuts implementing non-trivial 
monodromies. This structure occurs in the TQFT 
ground state(s) (the $(2$+$0)$-d case) as a result of 
the choice of holomorphic gauge. In the $(1$+$1)$-d case, 
however, it represents the actual critical dynamics 
of the excitations of the edge of the system. 
 
Although other possible settings are also suspected, 
there is only one place in nature where topological 
phases are known for certain to exist: the 
quantum Hall regime. In this regime, 
this `holographic' relationship between the $(2$+$1)$-d 
TQFT describing the bulk and the $(1$+$1)$-d RCFT 
describing edge excitations is beautifully realized. 
Since it is easier to experimentally probe the 
edge, the topological properties of quantum Hall states 
have, thus far, been mostly probed through 
the constraints that they place on the dynamics 
of the edge. 
 
One can continue to still lower dimensions 
and consider boundaries or defect lines 
in $2d$ classical critical systems and 
impurities in $(1$+$1)$-d quantum critical systems 
with dynamical critical exponent $z=1$. 
(In many $(2$+$1)$-d or $(3$+$1)$-d critical 
systems, only the $s$-wave channel interacts 
with the impurity. By performing a partial wave decomposition 
and keeping only this channel, one can map the 
problem to a $(1$+$1)$-d problem. Hence, the 
restriction to $(1$+$1)$-d is not very severe.) 
In these situations, the constraints of conformal 
invariance in the bulk strongly constrain the 
low-energy behavior of the boundary/defect 
correlations or impurity dynamics. 
The same methods can be applied to a point 
contact in a $(2$+$1)$-d system in a topological 
phase. The point contact allows tunneling 
between the gapless excitations 
at the edge and can, therefore, be understood 
as an impurity in a $(1$+$1)$-d system.  
 
In particular, we can write the entropy of the impurity problem as 
\begin{equation} 
\label{eqn:impurity-entropy-def} 
{\cal S}_{\hbox{imp}}= fL + \ln(g) \ +\ \dots 
\end{equation} 
 where $L$ is the size of one-dimensional space. The leading piece is
non-universal, but a universal part of the entanglement entropy
proportional to the central charge of the conformal field theory can
be extracted \cite{Casini04}. In this paper, we will focus on
subleading pieces of the entropy like $\ln(g)$.  Although $\ln(g)$ is
often called the ``boundary entropy'', we see that this interpretation
is not always useful: we will see how such a term can be present even
when there is no boundary to the $1$+$1$ dimensional system. Moreover,
if it were truly the entropy of the impurity, then $g$ would be an
integer. However, if one computes it by first taking the thermodynamic
limit $LT\to\infty$, and then taking $T\to 0$, $g$ is not always an
integer \cite{Andrei84,Tsvelik85,Cardy89,Affleck91}.  This makes this
subleading contribution to the entropy somewhat mysterious to
interpret physically.
 
To make this clearer, we will examine in detail the situation in which
the $(1$+$1)$-d critical system is the edge of a $(2$+$1)$-d system in
a chiral topological phase. In particular, we will show that the value
of $g$ in (\ref{eqn:impurity-entropy-def}) is closely related to that
of the quantum dimension ${\cal D}$ of the topological quantum phase,
that occurs in the entanglement entropy in
(\ref{eqn:topo-entanglement}). Although they are both subleading terms
in entropies, their relation is far from obvious. Indeed, their very
definitions are quite different, involving an interchanged order of
limits.  The entanglement entropy is a property of the ground
state obtained in the zero temperature limit, when the region
size $L$ is subsequently taken to infinity.  On the other hand,
extracting the subleading $\ln(g)$ correction of the thermodynamic
entropy as in (\ref{eqn:impurity-entropy-def}) requires taking $L$ to
infinity before taking the zero temperature limit.  Despite this
opposite ordering of limits, we will show that both quantities follow
from the same deep results in conformal field theory
\cite{Cardy89,Verlinde88,Moore88}.
 
The connection can be understood heuristically by considering the
entropy loss resulting from the impurity/point contact as the
temperature is decreased. This is convenient because it allows one to
get rid of the contributions proportional to $L$ by considering the
difference ${\cal S}_{UV}-{\cal S}_{IR}$, where ${\cal S}_{UV}$ is the
entropy in the absence of the impurity/point contact (the ultraviolet
limit), and ${\cal S}_{IR}$ is the entropy in the zero-temperature
limit (the infrared limit). In the IR limit, the point contact in the
topological theory effectively cuts the {\em edge} in two. When the
edge is cut into two, the topological entanglement entropy
(\ref{eqn:topo-entanglement}) receives a contribution
$-2\ln{\cal D}$. Thus ${\cal S}_{UV}-{\cal S}_{IR}=\ln{\cal D}$. As we
will see, by definition, ${\cal D}>1$, so this is indeed positive as
one expects for a flow caused by a relevant perturbation. This
demonstrates that a ``$g$-theorem'' \cite{Affleck91,Friedan04} applies
to a point contact across a system in a topological phase.  Likewise,
in quantum impurity problems the impurity degrees of freedom end up
being screened (at least partially) by coupling to the degrees of
freedom in the one-dimensional bulk.  

These results show that when a
$(1$+$1)$-d critical system is the boundary of a $(2$+$1)$-d system in
a topological phase, the thermodynamic ``boundary entropy'' of
eq. \ref{eqn:impurity-entropy-def} is actually the topological
entanglement entropy for the subsystems into which the system has been
dynamically split. This is reminiscent of the generation of entropy by
black hole formation. (The reader need not worry that by relating a
$(0$+$1)$-d entropy to a $(2$+$1)$-d entropy we are forgetting that
a boundary cannot have a boundary.  The point contact is not the
boundary of a $(1$+$1)$-d system, but rather a defect in the middle of
it -- and, therefore, capable of cutting it in two.)
 
In fact, the connection between the two kinds of entropy goes even 
deeper. When a topological phase has quasiparticles with non-Abelian 
statistics, there is an additional source of entanglement 
entropy. This arises because the Hilbert space for identical particles 
obeying non-Abelian statistics must be multi-dimensional (even when 
ignoring the position and momentum of the particles), so that the 
system moves around in this space as the quasiparticles are braided. 
For example, in a $p+ip$ superconductor, two vortices form a two-state 
quantum system, which therefore have an entanglement entropy 
$\ln(2)$. We dub this entropy the {\em bulk entanglement entropy}, 
because it comes from the gapped bulk quasiparticles. However, we 
emphasize that this a subleading piece of the 2$+$1-dimensional bulk 
entropy: it is independent of the size of the system, but rather 
depends only on the number and type of quasiparticle excitations in 
a given state. The remarkable connection is that this part of 2$+$1 
dimensional bulk entropy can be encoded in a 1$+$1 dimensional partition 
function. In other words, the entanglement entropy of these bulk 
quasiparticles is a property of the boundary of the system (just like 
a black hole!). We show how to compute such
{\em holographic partition functions} by using conformal field theory.

In section \ref{sec:quantum-dimensions}, we discuss the quantum 
dimensions of a topological state and of the various quasiparticle 
excitations of such a state. We give a general way to compute quantum 
dimensions using conformal field theory in section 
\ref{sec:qdfromcft}, and do this computation for a variety of examples 
in appendix \ref{sec:examples}. The examples include the Laughlin 
states for the abelian fractional quantum Hall effect, the 
Moore-Read\cite{Moore91} and Read-Rezayi\cite{Read99} non-abelian 
quantum Hall states, and the $p+ip$ 
superconductor\cite{Greiter92,Read00}.  In section \ref{sec:entropy} 
we derive the central results of our paper, which define the 
holographic partition function and how it describes the topological 
entanglement entropy.  We show in section \ref{sec:entropy-loss} how 
these results can be applied to understand how a point contact affects 
the entropy. This complements recent results of 
ours, relating point contacts in the Moore-Read state and in $p+ip$ 
superconductors to the Kondo problem \cite{Fendley06a,Fendley06b}.

\section{Quantum dimensions} 
\label{sec:quantum-dimensions} 
 
A fundamental characteristic of a topological field theory is the {\em 
quantum dimensions} of the excitations. The topological 
entanglement entropy of the systems studied here on a space with 
smooth boundaries is entirely given in terms of these numbers. 
 
For the quantum Hall effect and related 
states, one can find the quantum dimensions directly from the wavefunctions 
\cite{Nayak96c,Read96}. In this paper we find it useful to instead compute them 
using the methods of conformal field theory. The formal reason why 
this is possible is that the algebraic structure of rational conformal 
field theory is virtually identical to that of topological field 
theory (this can be understood precisely by using the mathematical 
language of category theory). A more intuitive reason is that the 
gapless edge modes of the 2$+$1-dimensional quantum theories studied 
here form an effectively 1$+$1 dimensional system. A 1$+$1-dimensional 
gapless system with linear dispersion has conformal invariance, so the 
powerful methods of conformal field theory are applicable. The edge 
modes are in one-to-one correspondence with the modes of the bulk 
system, so the $d_a$ computed using conformal field theory 
are those of the bulk system as well. 
 
We introduce the full conformal field theory formalism in the next 
section. In this section we explain how to define both the individual 
quantum dimensions of the quasiparticles $d_a$, and the total quantum 
dimension ${\cal D}$ of a theory. We discuss explicitly the 
simplest non-trivial example, the Ising model.

The quantum dimension is simple to define. 
Denote the number of linearly-independent states 
having ${\cal N}$ quasiparticles of type $a$ as $H_a({\cal N})$. Then the  
quantum dimension $d_a$ of the excitation of type $a$ is given by 
studying the behavior of $H_a({\cal N})$ for large ${\cal N}$, which behaves as 
$$H_a({\cal N})  
\propto d_a^{\cal N}\ .$$ To have non-abelian 
statistics, one must have $d_a>1$; these degenerate states are the 
ones which mix with each other under braiding. In such a 
situation one typically has $d_a$ which are not 
integers.

As we discussed in the introduction, the topological entanglement 
entropy is related to the total quantum dimension ${\cal 
D}$, which is defined as  
\begin{equation} 
{\cal D}=\sqrt{\sum_a d_a^2}. 
\label{totalD} 
\end{equation}  
The sum is over all the types of quasiparticles in the theory.  Thus 
to compute ${\cal D}$, one must not only compute the $d_a$ but be able 
to classify all the different quasiparticles as well. Since $d_a\ge 1$ 
by definition, for ${\cal D}$ to make sense there can only be a finite 
number of different quasiparticles. We must therefore confine 
ourselves to topological field theories where this is true. Luckily, 
the topological theories of most importance in condensed-matter 
physics have this property.

We will give a simple expression for ${\cal D}$ below in terms of the modular 
$S$ matrix of conformal field theory, but here we illustrate how it 
comes about.  In the fractional quantum Hall effect there is a very 
natural way of understanding the structure of the topological 
theory. Quasiparticle states are defined ``modulo an electron''.
This means that in the topological theory, two excitations 
which differ by adding or removing an electron are treated as 
equivalent. Thus the charge-$e$ electron is in the {\em identity} 
sector of the theory, since in the topological theory the identity and 
the electron are equivalent. 
 
A simple example is in the abelian Laughlin states at filling fraction 
$\nu =1/m$. We discuss in depth in the appendix how the consequence of 
modding out by an electron is that states differing in charge by any 
integer times $e$ are identified in the topological theory. Only 
fractionally-charged quasiparticles result in sectors other than the 
identity. In the Laughlin states, the fundamental quasiparticle has 
charge $-1/m$. One can obviously consider more than one of these to 
get larger fractional charge, but if one has $m$ of the fundamental 
quasiparticles, they can fuse into a hole (put another way, the hole 
can split apart into $m$ fundamental quasiparticles). Different 
sectors of the topological theory can therefore be labeled by the 
charges $ae/m$ where $a$ is an integer obeying $0\le a<m$. There are 
thus $m$ different quasiparticle sectors. The quantum dimension of any 
quasiparticle in an abelian theory is $1$, so we have $d_a=1$ for 
$a=0\dots m-1$. Using this with (\ref{totalD}) gives the total quantum 
dimension of a Laughlin state at $\nu =1/m$ to be 
\begin{equation} 
{\cal D}_{\nu=1/m} = \sqrt{m}. 
\end{equation} 
 
Finding ${\cal D}$ for non-abelian states requires knowing even more 
about the structure of the theory. Not only does it require knowing 
what the quasiparticles are, but finding the individual $d_a$ requires 
knowing what the {\em fusion rules} are. The fusion rules describe how 
to treat multi-particle states in the topological phase in terms of 
single-particle states. In the abelian case, the fusion rules are 
simple: for example, two charge-$e/m$ particles fuse to effectively 
give a single charge-$2e/m$ particle. The non-abelian structure of 
the theory arises when two particles can fuse in more than one way. 
 
We illustrate this in the $p+ip$ 
superconductor. As discussed in detail in a number of places \cite{Read00,Ivanov01},
a vortex has a Majorana fermion zero mode in its core. Then
two vortices share a Dirac fermion (=2 Majorana fermions)
zero mode. This zero mode can 
be either empty or filled; thereby, two vortices form a two-state 
system. Denoting the identity sector by $I$, the vortex by $\sigma$ 
and the filled zero mode by $\psi$, this means that the fusion rule 
for two vortices can be denoted by 
$\sigma\cdot\sigma = I + \psi.$ 
Two different terms on the right-hand-side means that there are two 
ways to fuse two sigma quasiparticles. (More precisely, this means 
that two vortices have a two-dimensional Hilbert space.) One finds 
that the full set of fusion rules here are 
\begin{eqnarray} 
\nonumber 
\psi\cdot\psi &=&I\\ 
\sigma\cdot\sigma&=& I\ +\ \psi \\ 
\label{Isingfusion} 
\psi\cdot\sigma&=& \sigma 
\nonumber 
\end{eqnarray} 
and, of course, fusion with the identity gives the same field back. 
 
The quantum dimensions follow from the fusion rules.  Two $\sigma$ 
quasiparticles here can fuse to give either the identity $I$, or the 
fermion $\psi$. Thus while $H_\sigma(1)=1$, these two possibilities 
for fusion of $\sigma$ with itself means that the dimension of the 
Hilbert space for two vortices is $H_\sigma(2)=2$.  Now 
include a third vortex. The dimension of the Hilbert space is not $4$, 
but rather is $H_\sigma(3)=2$. This is because 
$$\sigma\cdot\sigma\cdot \sigma = (I + \psi)\cdot\sigma = \sigma + \sigma.$$ 
Fusion is associative, so it makes no 
difference in which order we fuse. Continuing in this 
fashion, it is easy to see that  
$$H_\sigma(\cal N)= 
\begin{cases} 
2^{{\cal N}/2} \qquad \qquad&{\cal N} \hbox{ even}\\ 
2^{({\cal N}-1)/2} \qquad\quad& {\cal N} \hbox{ odd}\\ 
\end{cases} 
$$ The quantum dimension of the vortex is therefore 
$d_\sigma=\sqrt{2}$. Since $\psi\cdot\psi = I$, $H_\psi({\cal N})=1$ for all 
${\cal N}$, so the quantum dimension of the fermion $\psi$ is simply 
$d_\psi=1$. The quantum dimension of the identity field is obviously 
always $d_I=1$, so only the vortices exhibit non-abelian 
statistics. The total quantum dimension of the $p+ip$ superconductor 
is therefore 
\begin{equation} 
{\cal D}_{p+ip}=\sqrt{1^2 +1^2+ (\sqrt{2})^2}=2. 
\label{Dpp} 
\end{equation} 
As we will discuss in detail in the next section, this all follows 
from studying the Ising conformal field theory, which describes the 
edge modes of this superconductor.
 
\section{Computing the quantum dimensions using 
conformal field theory} 
 
\label{sec:qdfromcft} 
 
As discussed in the 
introduction, the bulk quasiparticles of a topological field theory 
are in one-to-one correspondence with the `primary 
fields' of a corresponding rational conformal field theory. Here we explain how to 
use this correspondence to give a systematic way to compute the 
quantum dimensions of the quasiparticles in a topological phase.

\subsection{Primary fields and quasiparticles} 
 
We start by reviewing some of the conformal field theory results we 
will be using. 
For an excellent introduction to much 
of this formalism, with detailed applications to the 
Ising model, see Ginsparg's lectures \cite{Ginsparg89}.

We take two-dimensional space to be a disk, so that its one-dimensional 
edge is a circle of radius $R$. At zero temperature the effective 
two-dimensional spacetime for the conformal field theory is  
therefore the surface of a cylinder.  We describe this cylinder by a periodic 
variable $0\le \theta < 2\pi$ and a (dimensionless) Euclidean time 
coordinate $\tau$.  It is often convenient to study the theory on the 
punctured plane with complex coordinates $z$ and $\overline z$, and 
then use the conformal transformations $z=e^{\tau+i\theta}$ and 
$\overline{z}=e^{\tau-i\theta}$ to map the results onto the 
cylinder. We will often study the theory at non-zero temperature 
$1/\beta$, so that $\tau$ becomes periodic with period $\beta$. At 
non-zero temperature, spacetime therefore is a torus. 
All the conformal field theories we study are {\em chiral}, 
which means that all fields depend only on $\tau+i\theta$ or $z$, and 
not $\tau-i\theta$ or $\overline{z}$. This is possible because of the 
time-reversal symmetry breaking of the 2$+$1 dimensional theory, coming, 
for example, from the magnetic field required for the Hall effect.

Many powerful techniques can be used to analyze conformal field theories. In two 
spacetime dimensions, conformal symmetry has an infinite number of 
generators. The symmetry is generated by the chiral part of the 
energy-momentum tensor $T(z)$ and its antichiral conjugate 
$\overline{T}(\overline{z})$. The generators of chiral conformal 
transformations are the modes of $T(z)$, i.e.\ the coefficients $L_n$ 
in the Laurent expansion $T(z) = \sum_n L_n z^{-n-2}$ on the punctured 
plane.  The energy-momentum tensor has dimension 2, so we have 
normalized the modes so that $L_n$ has dimension $n$. The Hamiltonian 
of the non-chiral system on the cylinder is 
$$H=\frac{2\pi}{R} (L_0+\overline{L}_0) - 
\frac{\pi c}{6R}.$$ The constant $c$ is known as the {\em central 
charge} of the conformal field theory.  The term proportional to $c$ 
arises from the conformal transformation of the plane to the cylinder; 
it can be interpreted as the ground-state or Casimir energy of the system 
in finite spatial volume $2\pi R$ \cite{BCNA}. Conformal symmetry 
requires that $[L_0,L_n] = -n L_n$, so acting on an energy eigenstate 
with $L_{-n}$ gives another eigenstate with energy shifted by $2\pi 
n/R$. Furthermore, an $L_0$ eigenstate has
eigenvalue which is equal to the dimension of the operator which creates it. 
It is convenient to define a ``chiral Hamiltonian'' by  
\begin{equation} 
{\cal H}= \frac{2\pi}{R} (L_0-c/24), 
\label{chiralH} 
\end{equation} 
so that $H ={\cal H} +\overline{\cal H}$. 
 
Since there are an infinite number of symmetry generators, the 
irreducible representations of conformal symmetry are 
infinite-dimensional. Thus one might hope to classify all the 
states of a 1$+$1-dimensional field theory in terms of a {\em finite} 
number of irreducible representations of conformal symmetry. In other 
words, there will be a finite number of highest-weight states, and all 
the other states of the theory will be obtained by acting on the 
highest-weight states with the $L_n$ with $n<0$. This is quite similar 
to decomposing states into irreducible representations of any 
non-abelian symmetry algebra such as, for instance, spin.
In fact, one often can extend 
the conformal symmetry algebra to a larger algebra by also including 
symmetries such as spin, supersymmetry, or even more exotic 
possibilities called $W$ algebras. One can then organize all the states 
of the theory into irreducible representations of these extended 
infinite-dimensional symmetry algebras. Theories with a finite number 
of highest-weight representations of such extended symmetry algebras 
are called {\em rational conformal field theories}. The fields 
creating the highest-weight states are called {\em primary fields}, 
and those obtained by acting on these with the symmetry generators are 
called {\em descendants}. 
 
For the theories of interest here, there are in general an infinite 
number of primary fields under the conformal symmetry, but a finite 
number under a larger symmetry group. There is a marvelous physical 
interpretation of the presence of this extended symmetry algebra. We 
discussed in the previous section how in the Hall effect, the 
topological field theory arises by considering all the states ``modulo an 
electron''. In the edge conformal field theory, the symmetry
algebra can be extended by including the electron 
annihilation/creation operators.  
Thus {\em  the primary fields of the conformal field theory are in 
  one-to-one correspondence with the quasiparticles in the topological 
  phase.}  
Descendant states arise by attaching electrons or holes to the quasiparticles -- acting 
with the electron creating or annihilation operator moves one around 
inside an irreducible representation of the extended symmetry 
algebra. We will give explicit examples below of how this works.


\subsection{Quantum dimensions from the fusion rules}

There are several equivalent ways of using conformal field theory to 
compute the quantum dimensions of the quasiparticles. All amount to 
finding the {\em fusion coefficients}. Since all the fields of the 
theory are expressed by acting with the symmetry generators on the 
primary fields, the operator-product expansion of two primary fields 
can be written as a sum over the primary 
fields. Namely, the ``fusion rules'' of a 
conformal field theory are written as 
\begin{equation} 
\phi_a \cdot \phi_b = \sum_c N_{ab}^c \phi_c 
\label{fusion} 
\end{equation} 
where the fusion coefficients 
$N_{ab}^c$ are non-negative integers, which count the number of times 
the primary field $\phi_c$ appears in the operator-product expansion of 
$\phi_a$ and $\phi_b$. This algebra is associative, and one has 
$N_{ab}^c=N_{ba}^c=N_{bc}^a$. For all 
the theories considered here, we have $N_{ab}^c=0$ or $1$, but it is 
possible to have larger integers in more complicated theories.

Each quasiparticle in the topological field theory corresponds to a 
primary field, and the fusion of these quasiparticles is the same as 
that of the corresponding primary fields. Thus to have non-abelian 
statistics, $N_{ab}^c$ for some $a$ and $b$ must be non-zero for 
more than one $c$. 
A simple example of non-abelian statistics is given by the Ising
model, which has three primary fields: the identity field $I$, the
fermion $\psi$, and the spin field $\sigma$ which have the fusion
rules given in (\ref{Isingfusion}) above. The Ising CFT describes
the edge excitations of a $p+ip$ superconductor, with the vortices
corresponding to $\sigma$.
\cite{Read00,Ivanov01}  This can be shown, for example, by using the
Bogoliubov-de Gennes equations for the superconductor, as reviewed in
ref.\ \onlinecite{Fendley06b}. The non-zero fusion
coefficients are
$N_{\sigma\sigma}^I=N_{\sigma\sigma}^\psi=N_{\sigma\psi}^\sigma =
N_{\psi\psi}^I = 1$.  These fusion rules are compatible with the
spin-flip symmetry $\sigma\to -\sigma$, $\psi\to \psi$.
 
We detailed in the previous section how to compute the quantum 
dimensions for the Ising fusion rules. The general procedure is 
similar. If we fuse $M$ $\phi_a$ fields together and use
(\ref{fusion}) recursively on the right-hand side:
\begin{eqnarray} 
\phi_a \cdot \phi_a \cdot \ldots \cdot \phi_a = 
N_{aa}^{c_1} N_{ac_1}^{c_2} \cdots N_{ac^{}_{M-2}}^{c_{M-1}} \phi_{c^{}_{M-1}} 
\end{eqnarray}
This is the product of $M-1$ copies of the matrix 
\begin{equation} 
(Q_a)^c_b = N_{ab}^c \ .
\label{QN} 
\end{equation}
In the $M\rightarrow\infty$ limit, the product will be dominated by the
largest eigenvalue of $Q_a$. This is the quantum dimension $d_a$.
 
This eigenvalue can easily be written in terms of the {\em modular S 
 matrix}, which we will define and discuss in the next subsection 
 \ref{modS}. A profound result of conformal field theory is the {\em 
 Verlinde formula}\cite{Verlinde88,Moore88}, which expresses the fusion rules in terms of the 
 elements of $S$, namely, 
\begin{equation} 
N_{ab}^c = \sum_j \frac{S_a^j S_b^j S^c_j}{S_0^j}  
\label{verlinde} 
\end{equation} 
where $0$ denotes the identity field and the sum is over all 
primaries. In general $S$ is unitary and hermitian; 
in all of our examples $S$ is also real and hence 
symmetric. We therefore give the formulas for this case; they are simple 
to generalize. 
 
To find the eigenvalues and eigenvectors of $Q_a$, we 
multiply the Verlinde formula by $S^b_k/S_0^k$ and 
sum over $b$. This yields 
$$\sum_{b} (Q_a)^c_b \frac{S^b_k}{S_0^k} =  
\frac{S_a^k}{S_0^k} \frac{S^c_k}{S_0^c}\ .$$  
The $b$th element of the eigenvector of $Q_a$ with eigenvalue 
$S_a^k/S_0^k$ is therefore given by $S^b_k/S_0^k$.   
The modular $S$ matrix is the matrix of the 
eigenvectors of $Q^a$. The 
largest eigenvalue is the one with $k=0$, so we have 
\begin{equation} 
d_a = \frac{S_a^0}{S_0^0}. 
\label{quantdim} 
\end{equation} 
 
There is a simple expression for the total quantum dimension ${\cal 
D}$ in terms of the modular $S$ matrix.  
Using 
(\ref{quantdim}) along with the fact that $S$ is unitary and symmetric gives 
\begin{equation} 
{\cal D} = \sqrt{\sum_a \left(\frac{S_a^0}{S_0^0}\right)^2} =
\frac{1}{S_0^0} \ .
\label{DS} 
\end{equation} 
This formula will prove useful in relating the topological and 
thermodynamic entropies.

\subsection{The modular $S$ matrix} 
\label{modS}

Rational conformal field theories have a variety of profound 
mathematical properties. We have exploited one of them in using  
the Verlinde formula (\ref{verlinde}) to give a compact expression 
(\ref{quantdim}) for the quantum dimensions in terms of the modular 
$S$ matrix. In this section  
we will discuss what $S$ is and how to compute it. 
 
To define $S$ and to use the Verlinde formula, we study conformal field 
theory on a torus, i.e.\ in finite size in both the 
Euclidean time and spatial directions. Physically, this 
corresponds to a non-zero temperature, so that the Euclidean time 
coordinate is periodic with period $\beta$. One must  
compute the partition function  
in the sector of states associated with each primary field $a$ and its 
descendants. In mathematical language, 
this is called a character, and is defined as 
\begin{equation} 
\chi_a(q) \equiv \hbox{ tr}_a e^{-\beta {\cal H}}  
= q^{-c/24}\hbox{ tr}_a q^{L_0}, 
\label{chia} 
\end{equation} 
where $q\equiv e^{-2\pi \beta/R}$, $R$ is the spatial length of
the system, and the trace $\hbox{ tr}_a$ 
is over all the states in 
the irreducible representation of the extended symmetry algebra 
corresponding to the primary field $a$ (i.e.\ the highest-weight state 
and its descendants). The partition function of the chiral theory is 
then of the form 
\begin{equation} 
Z = \sum_a N_a \chi_a(q), 
\label{ZN} 
\end{equation} 
where the $N_a$ are integers, representing how many copies of each 
primary field appears in a given theory. In a 
non-chiral theory, one defines analogous antichiral partition 
functions $\overline{\chi}_a(\overline{q})$, and the full partition 
function of a rational conformal field theory is given by a sum over 
products of the form $\chi_a(q) \overline\chi_b(\overline{q})$, 
again with integer coefficients. In both chiral and non-chiral 
rational conformal field theories, the sums are over a finite number 
of characters. We will give 
explicit examples of these characters below. 
 
The modular $S$ matrix describes 
how the characters behave when one exchanges
the roles of space and time:
\begin{equation} 
\chi_a(\widetilde{q}) = \sum_b S_{a}^b \chi_b(q). 
\label{Sdef} 
\end{equation}
where $\widetilde{q}\equiv e^{-2\pi R/\beta}$.
Keep in mind, however, that a $1$+$1$-dimensional quantum theory 
in Euclidean time is equivalent to a two-dimensional classical 
theory. Going back from the two-dimensional classical theory
to a $1$+$1$-dimensional quantum theory, one is free to choose 
either of the directions to be space and the other to be time.
We thus conclude that the theory should be invariant under
modular transformations of the torus. Hence, 
\begin{equation}
Z = \sum_a \widetilde{N}_a \chi_a(\widetilde{q})\ .
\end{equation} 
However, the $\widetilde{N}_a=\sum_b S^b_a N_b$ are usually not 
integers; we will have much more to say about this in section 
\ref{sec:entropy}.

Computing the characters is a problem in the representation theory of 
the extended symmetry algebra. In the cases of interest here 
the computation is straightforward; we give several examples in 
appendix \ref{sec:examples}. Finding the integers $N_a$ for a given bulk topological 
state is the crucial computation in this paper, and we discuss how 
these arise, and what they have to do with the entropy, in section 
\ref{sec:entropy}.

\section{Holographic partition functions} 
\label{sec:entropy} 
 
In this section we arrive at a central result of this paper. We 
first define the edge entropy in terms of the chiral conformal field 
theory describing the edge modes. The edge entropy is not merely 
reminiscent of the topological entanglement entropy  
(\ref{eqn:topo-entanglement}) discussed in the introduction, but 
has an identical universal part $-\ln({\cal D})$.
We then extend the correspondence between entanglement entropy 
and thermodynamic entropy by studying the bulk entanglement 
entropy, which arises from the different fusion channels of the bulk 
quasiparticles. We show that both of these entropies can be encoded in 
a single holographic partition function, which we interpret as the 
topological entanglement entropy of a region of a system in a 
topological phase.

\subsection{Edge and ``boundary'' entropies} 
 
We are studying 2$+$1 dimensional systems which are gapped
in the bulk but gapless at the edge. Their edge 
modes are described by {\em chiral} rational conformal field 
theories. In the previous sections, we have defined and discussed the 
characters of conformal field theories, which are essentially chiral 
partition functions. Here we use these to characters to define the 
{\em edge entropy}.

\begin{figure}[h] 
\begin{center} 
\includegraphics[width= .4\textwidth]{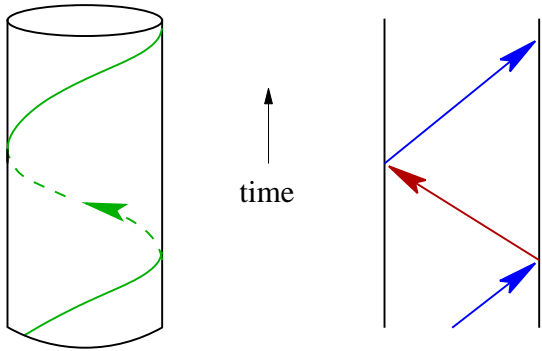} 
\caption{The edge modes with spacetime a cylinder (illustrated on the
left) have only one chirality. The physics of these is closely
related to that of a non-chiral system with spacetime a strip
(illustrated on the right), where 
the boundary conditions couple the left and right movers.} 
\label{fig:fig1} 
\end{center} 
\end{figure}
To motivate our definition of the entropy, we first discuss a 
closely-related system, a non-chiral rational conformal field theory with 
boundaries. This means that spacetime at zero temperature is 
a strip of width $\pi R$ instead of a cylinder. At 
non-zero temperature, spacetime becomes a finite-width cylinder 
with period $\beta$ in the Euclidean time direction, instead of a torus. 
The left and right movers on the strip are 
coupled by the boundary conditions, as illustrated schematically in
figure \ref{fig:fig1}. This means the chiral and antichiral 
conformal symmetries are also coupled. Conformal invariance is 
preserved only with certain boundary conditions, and even then, only a 
single algebra remains \cite{Cardy84}. This symmetry algebra is 
exactly the same as that of a {\em single} chiral theory.  The 
partition function for the model on the strip is then a sum over 
characters of a single algebra. These characters are exactly those in 
(\ref{chia}), and the partition function takes the same form 
(\ref{ZN}), with $N_a$ integer \cite{Cardy86b,Cardy89}. 
  
The connections between the chiral theory on the cylinder and the
non-chiral theory on the strip go even deeper. The boundary conditions
which preserve conformal invariance are in one-to-one correspondence
with the primary fields of this chiral rational conformal field
theory, so each can be labeled by the same indices we use to label
primary fields. (More precisely, conformal boundary conditions form a
vector space, basis vectors of which can be labeled by the primary
fields.) Then a key result of Cardy's is that with
appropriate regularization, the partition function $Z_{jk}$ on the
strip with boundary conditions $j$ and $k$ at the two ends is \cite{Cardy89}
\begin{equation} 
Z_{jk} = \sum_a N^a_{jk} \chi_a(q) 
\end{equation} 
where $N^a_{jk}$ is the {\em same} $N^a_{jk}$ which appears in the 
fusion rules! Furthermore, by using the Verlinde 
formula it is simple to find boundary conditions which result in a 
partition function given by a single character $\chi_a$.  
 
These results were used by Affleck and Ludwig to define and compute 
what they called the ``boundary entropy''\cite{Affleck91}. Consider an 
RCFT on the finite-width cylinder with boundary conditions $j$ and 
$k$, so that the partition function is $Z_{jk}$. Then take the width 
of the cylinder $R\to\infty$ while holding the temperature, and hence the 
radius $\beta$ of the cylinder, fixed at some non-zero finite 
value. Then, on general grounds, one must have 
\begin{equation} 
\ln{Z_{jk}} = fR + \ln(g) + \dots 
\label{Zfg} 
\end{equation} 
where we neglect terms that vanish as $R\to\infty$. This is precisely 
the form (\ref{eqn:impurity-entropy-def}).
We can compute $f$ 
and $g$ by using a modular transformation, because the limit 
$R/\beta\to\infty$ corresponds to $\widetilde{q}=0$. {}From 
(\ref{Sdef}) we have 
$$Z_{jk} = \sum_{a,b} N^a_{jk} S_a^b\chi_b(\widetilde{q}).$$ 
By definition, in this limit, 
the character $\chi_a$ behaves as  
$$\lim_{\widetilde{q}\to 0} \ln \chi_a = \frac{2\pi R}{\beta} \left(\frac{c}{24} - h_a\right)$$ 
So unless $j$ and $k$ are such 
that $\sum_a N^a_{jk} S_a^0=0$, the identity character $\chi_0^{}$ 
dominates and one has 
\begin{equation}  
f = \frac{\pi c}{12 \beta}, \quad\qquad g=\sum_a N^a_{jk} S_a^0. 
\label{fg} 
\end{equation} 
It is then natural to interpret $g$ as a ground-state degeneracy, 
and $\ln(g)$ as a boundary entropy, since clearly this subleading term 
in the full entropy depends on the boundary conditions $j$ and 
$k$. However, $g$ is not necessarily an integer, so as we discussed in 
the introduction, it cannot, strictly speaking, be a ground-state degeneracy. 
Neither is it solely associated with the boundaries of the strip,
since there is no particular reason that subleading terms 
from the 1$+$1 dimensional bulk cannot contribute to $\ln(g)$. 
 
We now return to studying topological theories whose edge modes are 
described by chiral conformal field theories on the torus. Although 
the torus has no boundaries, there are still boundary conditions on 
the fields. For example, in the $p+ip$ superconductor, the edge 
fermions have antiperiodic or periodic boundary conditions around the spatial 
cycle of the torus depending on whether there are an even or an odd 
number of vortices in the two-dimensional bulk. So let us first 
consider the case where there are no bulk quasiparticles. In the 
language of topological field theory, this corresponds to trivial 
topological charge on the disk. The partition function of the edge 
RCFT is then 
$$Z_0 = \chi_0^{}(q)$$ where, as above, the $0$ label on $\chi_0^{}$ means the 
identity sector.  To extract the piece of interest from this, we take 
the same $R/\beta\to\infty$ limit. $Z_0$ then can 
be expanded in the form (\ref{Zfg}), i.e. 
\begin{equation}
\ln(Z_0)= f_0 R + \ln(\gamma_0),
\label{Zgamma}
\end{equation} where $\gamma$ in this chiral theory is the 
analog of the $g$ in (\ref{Zfg}) in the impurity theory. Following
through the same modular transformation described in the preceding
paragraph, we obtain the subleading contribution to $\ln(\chi_0^{})$
in this limit:
$$\gamma_0 = N_{00}^0 S_0^0 = S_0^0.$$ Recall, however, that we showed in 
(\ref{DS}) that $S_0^0=1/{\cal D}$, where ${\cal D}$ is the quantum 
dimension we have gone to great lengths to compute. We thus conclude 
\begin{equation} 
{\cal S}_{\hbox{edge}}\equiv \ln(\gamma_0) = -\ln({\cal D}) 
\label{Sedge} 
\end{equation} 
as the universal part of the {\em edge entropy}. There is no 
particular reason for $\gamma_0$ to be an integer, since it is a subleading 
term in the large $R/\beta$ expansion. 
 
Note that ${\cal S}_{\hbox{edge}}$ is identical to the universal piece of the 
topological entanglement entropy in 
(\ref{eqn:topo-entanglement}). Although the two quantities are 
defined in different ways, this is hardly a 
coincidence. ${\cal S}_{\hbox{edge}}$ is defined for a system with a real 
edge, while the topological entanglement entropy is defined on the 
boundary between two regions $A$ and $B$, not a physical 
edge. Nevertheless, one expects the two to be the same. Edge modes 
arise formally in a topological theory to cancel the chiral anomaly 
\cite{Wen91}. The topological entanglement entropy arises formally by 
integrating out the system beyond the boundary, i.e.\ subsystem 
$B$. When integrating out the degrees of freedom of $B$,
one of course integrates out an anomaly-free theory,
so this must include the edge modes of $B$ as 
well. The remaining $A$ theory then, in some sense, must  include ``edge modes" as well
which are required to cancel those on $B$.
This required cancellation is one way of seeing 
that systems $A$ and $B$ are indeed entangled. Thus, the 
value of the topological entanglement entropy should be 
identical to our computation of the edge entropy. We have shown by 
direct computation that this is indeed the case.

\subsection{Bulk entanglement entropy} 
 
The correspondence between thermodynamic and
entanglement entropies goes much deeper. Let us now consider the
{\em bulk entanglement entropy}. This arises in non-abelian topological phases because of 
multiple possible fusion channels. For example, two vortices in a $p+ip$ 
superconductor form a two-state quantum system, so their entanglement 
entropy is $\ln(2)$.  
 
We can derive a simple formula for the bulk entanglement entropy in 
general. We define the {\em topological degeneracy} of a given state 
to be the number of ways one can fuse the quasiparticles present to 
get a particular overall fusion channel. This is easily written in 
terms of the matrix $Q_a$ defined in (\ref{QN}). When there are $n_a$
of each of the quasiparticles labeled by $a$, define the matrix $T$ by
\begin{equation} 
T=\prod_a (Q_a)^{n_a} 
\label{TQ} 
\end{equation} 
It makes no difference in which order we multiply the $Q_a$ 
because different $Q_a$ commute with each other (as we 
showed from the Verlinde formula, they have the same eigenvectors). 
The topological degeneracy of channel $c$ is then
the $c,0$ entry of this matrix $T$. 
Note that this is a simple generalization of the way we computed the 
quantum dimension to the case in which different types of 
particles are allowed to be present.

Let us illustrate this in terms of the $p+ip$ superconductor, where 
there are three states in the topological field theory, labeled by 
$I$, $\psi$ and $\sigma$.  If there are no bulk quasiparticles 
present, we have degeneracy $1$ in the identity channel.  If we have 
only $\psi$ quasiparticles, the fusing is simple: for an even number 
of $\psi$ quasiparticles the degeneracy is $1$ in the identity channel, 
while for an odd number the degeneracy is $1$ in the $\psi$ channel. 
It gets more interesting for bulk $\sigma$ quasiparticles. A single 
bulk $\sigma$ quasiparticle corresponds to degeneracy $1$ in the 
$\sigma$ channel. However, two bulk $\sigma$ quasiparticles does {\em 
not} correspond to degeneracy $2$. Since $\sigma\cdot\sigma =I + 
\psi$, two bulk $\sigma$ quasiparticles corresponds to degeneracy $1$ 
in both the $I$ channel and in the $\psi$ channel. Degeneracy $2$ in 
the $\sigma$ channel occurs for {\em three} bulk $\sigma$ 
quasiparticles. In general, for an even number $2M$ of bulk $\sigma$ 
quasiparticles, we have degeneracy $2^{M-1}$ in the $I$ and $\psi$ channels 
while for an odd number $2M+1$, we have degeneracy $2^{M}$ in the 
$\sigma$ channel. Including any number of $\psi$ quasiparticles doesn't 
change the degeneracies (except for $M=0$). 
 
The bulk entanglement entropy arises from the 
uncertainty in knowing which quantum state the bulk quasiparticles are 
in. For example, if there are two bulk $\sigma$ quasiparticles, we do 
not know without doing a measurement whether they are in the $I$ or $\psi$ 
channel. The entropy is therefore $\ln(2)$. But what if there is just 
one $\sigma$ quasiparticle in a region? It can be entangled with a 
$\sigma$ quasiparticle in another region of the droplet, giving a 
total entropy of $\ln(2)$. Since entropy is additive, the only 
consistent possibility is for the single $\sigma$ quasiparticle to 
have entropy $\ln(\sqrt{2})$. In general, a single quasiparticle of 
type $a$ has entropy $\ln(d_a)$, where $d_a$ is the quantum dimension 
discussed at great length above.  
The general expression for the bulk entropy 
can be expressed easily in terms of 
the topological degeneracies, namely 
\begin{equation} 
{\cal S}^{}_{\hbox{bulk}}= \ln\!\biggl(\sum_c T_{0c} {d_c}\biggr) \ .
\label{SbulkT} 
\end{equation} 
This can be simplified by rewriting $d_c$ and $T_{c0}$ in terms of the 
modular $S$ matrix. We can use the Verlinde formula (\ref{verlinde}) for 
$Q$ and the relation $d_c=S_c^0/S_0^0$. The latter is the 
eigenvector of any $Q_a$, having eigenvalue $S^0_a/S_0^0=d_a$.
Thus it is an eigenvector of $T$ as well, so
\begin{equation} 
{\cal S}^{}_{\hbox{bulk}}= \sum_a n_a\ln(d_a)\ . 
\label{Sbulk} 
\end{equation} 
Each quasiparticle of quantum dimension $d_a$ contributes $\ln(d_a)$ 
to the bulk entanglement entropy. Note that this ``bulk'' entropy is 
extensive not in the size of the sample, but rather in the number of 
bulk quasiparticles.

\subsection{The bulk and edge entropies from the  
holographic partition functions} 
 
We show here that the {\em entire} topological entanglement 
entropy can be seen in terms of the edge conformal field theory. We 
interpret this as an example of {\em holography}. In a topological 
field theory, the degrees of freedom live on the edge. In the 
fractional quantum Hall effect, these are, of course, the edge modes we 
have been discussing all along. Holography simply means that all the 
information of the bulk topological field theory is encoded in the 
edge modes. 
 
As we have discussed above, adding or removing bulk quasiparticles 
changes the boundary conditions on the edge modes. This will change 
the chiral partition function for the edge. To illustrate this, 
consider the $p+ip$ superconductor, where there are 
three characters: $\chi_I$, $\chi_\psi$ and 
$\chi_\sigma$, with a modular $S$ matrix given in (\ref{SIsing}).  
The presence of $\sigma$ quasiparticles (the 
vortices) in the bulk has an important effect on the 
edge. As we reviewed in ref.\ \onlinecite{Fendley06b}, an odd number of $\sigma$ 
quasiparticles changes the boundary conditions of the edge Majorana 
fermion from antiperiodic around the circle to periodic. For a fermion 
with periodic boundary conditions on the 1$+$1 dimensional torus, the 
chiral partition function is $\chi_\sigma$. Thus we define the holographic partition function 
for a {\em single} bulk $\sigma$ quasiparticle to be 
$Z_{\sigma}=\chi_\sigma.$  
This partition function yields the correct entanglement entropy. 
Taking the $L/\beta\to\infty$ limit as in (\ref{Zgamma}) yields 
$$\gamma_\sigma= S_I^\sigma = \frac{1}{\sqrt{2}}.$$ 
For a $p+ip$ superconductor with one vortex in region $A$, the 
entanglement entropy is indeed 
$${\cal S}_{\hbox{bulk}}^{}+{\cal S}_{\hbox{edge}}^{}= \ln(\sqrt{2}) - 
\ln(2)=-\ln(\sqrt{2}).$$ We define the holographic partition function 
for a single bulk $\psi$ quasiparticle in a similar fashion: 
$Z_\psi=\chi_\psi$. The corresponding entropy is then 
$\ln(S^\psi_I)=\ln(1/2)=-\ln{\cal D}$. This is just ${\cal S}_{\hbox{edge}}$, 
which is indeed what we expect, because there is no bulk topological 
entropy for $\psi$ quasiparticles.

Given our definition of the bulk entanglement entropy, it should be 
obvious how to define the holographic partition function for an 
arbitrary number of bulk particles. The effect of adding a bulk 
quasiparticle is given by the same fusion rules as for the primary 
fields. For example, in the $p+ip$ superconductor, the fusion rule 
$\psi\cdot\psi=I$ means that with no $\sigma$ 
quasiparticles and an even (odd) number of $\psi$ quasiparticles, we 
define the holographic partition function to be $\chi_I$ 
($\chi_\psi$). In both cases the entropy remains $-\ln{\cal D}$, 
since there is no bulk entanglement entropy. A less trivial case is 
when we have two bulk $\sigma$ quasiparticles. Here the holographic partition 
function follows from the fusion rule $\sigma\cdot\sigma = I +\psi$, 
so we get  
$$Z_{2\sigma}=\chi_I + \chi_{\psi}.$$ 
For three, the topological degeneracy is $2$ in the $\sigma$ channel, 
so we get 
$$Z_{3\sigma}=2\chi_{\sigma},$$  
and so on. 
 
The definition of the holographic partition function for arbitrary 
numbers of quasiparticles in the bulk should now be obvious. 
In general, for $n_a$ bulk quasiparticles of each type $a$, the 
holographic partition function is 
\begin{equation} 
Z = \sum_a T_{0a} \chi_a(q) 
\label{ZT} 
\end{equation} 
where $T$ is the topological degeneracy matrix defined in (\ref{TQ}).  
This is the partition function obtained by computing the path
integral for the appropriate 2$+$1 dimensional topological field theory
(e.g.\ Chern-Simons theory) with spacetime $D\times S_1$, where $D$ is
a disk of radius $R$, and $S_1$ a circle of radius $\beta$. To get the
correct degeneracies $T_{0a}$ one needs to insert $n_a$ Wilson loops
of each type $a$ which puncture the disk and wrap around the $S_1$.

As advertised, the holographic partition function (\ref{ZT}) gives the
correct entanglement entropy in the the $R/\beta\to\infty$ limit. In
this limit,
$$ 
Z= \sum_a T_{0a}S_a^0 \chi_0^{}(\widetilde{q}) +\dots\ .
$$ 
We define the total thermodynamic entropy ${\cal S}$
via our usual expansion
$$\ln(Z) = fR + {\cal S} + \ldots\ .$$ 
Using $d_b=S^b_0/S_0^0$ along with the expressions for the bulk and edge 
entropies (\ref{SbulkT}) and  (\ref{Sbulk}) gives 
\begin{eqnarray} 
\nonumber 
{\cal S}&=& \ln (\sum_a T_{0a}S_a^0)\\ 
\nonumber 
&=& \ln (\sum_a T_{0a} d_a) + \ln(S_0^0)\\ 
\label{Stot} 
&=&\sum_a n_a  \ln (d_a) - \ln({\cal D})\ . 
\end{eqnarray} 
Thus we see that indeed once we have taken the $R/\beta\to\infty$
limit, we have 
\begin{equation} 
{\cal S}={\cal S}_{\hbox{bulk}} + {\cal S}_{\hbox{edge}} \ .
\label{SSS} 
\end{equation} 
The holographic partition includes both bulk and edge entanglement 
entropies.
 
We close this section by returning to where it started, a discussion of 
the relation between chiral conformal field theory with 
spacetime a torus and non-chiral conformal field theory on a finite-width 
cylinder. The partition functions in both cases are expressed in the form 
$$Z=\sum_a N_a \chi_a $$  
where the $N_a$ are integers, describing how 
many times each character appears. In the latter case, however, there 
is one important restriction which we need not demand of the holographic 
partition function. For a partition function describing a unitary 
field theory on the finite-width strip with physical boundary 
conditions, the identity character usually appears at most once (i.e.\ 
$N_0=0$ or $1$). The reason 
is that there is only one identity field in a unitary theory, so only
with peculiar boundary conditions can one have $N_0>1$. In the 
holographic partition function, $N_0$ can be any non-negative 
integer: it is the number of ways the bulk quasiparticles  
in the 2$+$1 dimensional topological theory can fuse to 
get the identity channel.  Thus although the partition functions are 
defined in a very similar fashion,  they really are different objects.

\section{Dynamical Entropy Loss at a Point Contact} 
\label{sec:entropy-loss} 

The relation between the thermodynamic and
entanglement entropies manifests itself most dramatically
through the dynamics of a point contact connecting two points on the
boundary of the disk. Quasiparticles can tunnel between
the two points connected by the contact, thereby perturbing
the edge modes at these two points. This can change the edge entropy.
When the tunneling operator is a relevant perturbation,
it causes the system to flow to an infrared fixed point at which
any edge excitation which is incident upon the contact
necessarily tunnels across the system. In other words, an imaginary 
boundary between the two halves evolves dynamically into a real edge.
As we shall see below, this process causes the entanglement entropy
between the two halves to be, in a sense, ``reborn" as the actual thermodynamic
entropy of the two resulting droplets.
We have discussed a point contact in a non-abelian topological state
in depth in a companion paper \cite{Fendley06b}, and the results in this section 
complement those results. 

As mentioned in the introduction, if we compute the
density matrix for half of a system by tracing out the degrees
of freedom of the other half, the resulting density matrix
has an entanglement entropy
$$ 
{{\cal S}_A} = aL - \ln {\cal D} + \ldots 
$$
The second, universal term is negative. What it tells us is that
there is a little less uncertainty in the state of half of the system
than one might have naively expected purely from local
correlations\cite{Kitaev05,Levin05}. {\it A priori}, this has absolutely nothing
to do with the thermodynamic entropy of the system since
it is obtained from an arbitrary division of the ground state
wavefunction alone, with nothing about the spectrum
taken into account. However, when the strong tunneling fixed
point is reached, the disk is effectively split into two.
The imaginary division into two halves becomes a real
division. Entropy is lost as a result of this process
because the edge entropy for two disks is $-2\ln({\cal D})$,
compared to $-\ln({\cal D})$ for a single disk.
As in the case of the entanglement entropy, this extra negative
contribution is due to the fact that we know more
about the system: the negative contribution $-\ln({\cal D})$
to the entropy of a disk represents the information that we know about
the system when we know that it has trivial total topological charge.
However, one half of the system could have arbitrary
topological charge $a$, so long as the other half
has the compensating topological charge $\bar{a}$.
If the point contact breaks the system into two
systems, each with trivial topological charge,
then this uncertainty is lost. The information which we now know
about the two disks is $-2\ln({\cal D})$.

Defining ${\cal S}_{UV}$ to be the entropy in the limit in which there is 
no tunneling across the sample (the UV limit), and ${\cal S}_{IR}$ to be the 
entropy in the limit in which tunneling is strong (the IR limit), we have 
\begin{equation} 
{\cal S}_{UV}-{\cal S}_{IR} = -\ln ({\cal D}) - (-2\ln({\cal D})) = 
\ln({\cal D})\ .
\label{SUVIR} 
\end{equation}
Here, we have assumed that there are no quasiparticles in the bulk,
and that the system breaks into two halves each with no quasiparticles.
However, the entropy change is the same even when quasiparticles are
present, which we will discuss below.
 
Let us first illustrate entropy loss due to a point contact in the
abelian case, when the edge theory consists of a free boson. This has
been widely studied in the context of the Laughlin fractional quantum
Hall states \cite{Kane92}. By using a series of mappings, the chiral
problem on the torus can be shown to be exactly equivalent to a
non-chiral problem on the finite-width cylinder. In other words, the
analogy we made in the previous section is an exact equivalence in
this case. The point contact results in a boundary with Neumann
boundary conditions on the boson in the UV limit, and Dirichlet
boundary conditions in the IR \cite{Wong94}.  The edge entropies can
be computed in both cases, and one obtains \cite{Wong94,Fendley94} for
$\nu=1/m$
$${\cal S}_{UV}-{\cal S}_{IR} = \ln(\sqrt{m}).$$ This is exactly what
one obtains from our results by assuming that the IR limit consists of
two separate disks. We show in (\ref{DLaughlin}) that ${\cal
D}=\sqrt{m}$ for the $\nu =1/m$ Laughlin states. Thus we indeed have
${\cal S}_{UV}-{\cal S}_{IR} = \ln({\cal D})$ as in (\ref{SUVIR}).
 
Using the results derived above, we can check (\ref{SUVIR}) in several 
non-abelian cases. As we showed in ref.\ \onlinecite{Fendley06b}, a 
point contact in the $p+ip$ superconductor is equivalent to the 
single-channel anisotropic Kondo problem. The entropy change is 
well-known here. The Kondo problem consists of a single impurity spin 
antiferromagnetically coupled to an electron gas. In the 
single-channel spin-1/2 case, the spin is decoupled in the UV 
limit, while it is completely screened in the IR limit. Therefore 
${\cal S}_{UV}=\ln(2)$, while ${\cal S}_{IR}=\ln(1)=0$. We showed in (\ref{Dpp}) 
that for the Ising model, ${\cal D}_{p+ip}=2$, so (\ref{SUVIR}) holds.

A point contact in the Moore-Read Pfaffian state at $\nu=1/2$
(or $\nu=5/2$) turns out to be a 
variant of the two-channel overscreened Kondo problem.  In ref.\ 
\onlinecite{Fendley06a}, we showed that the entropy loss here is 
${\cal S}_{UV}-{\cal S}_{IR}=2\ln(\sqrt{2})$.  In (\ref{DMR}), we show that ${\cal 
D}_{MR}=2\sqrt{2}$, so again (\ref{SUVIR}) holds. For a point contact in 
a Read-Rezayi state or for $SU(2)_k$ for general $k$, it is not yet 
known how to map the tunneling operator onto a Kondo-like problem, so
we do not have an independent result against which to
check the prediction which can be obtained from 
(\ref{SUVIR}) along with the quantum dimension in (\ref{DRR}).

The entropy change (\ref{SUVIR}) can also be seen in terms of the 
holographic partition functions.  The partition function for  
two independent systems is simply the product of the partition 
functions for the two systems. Thus when the point contact splits the 
system into two, the holographic partition function should be the product 
of two holographic partition functions. Let us first examine this in 
the case where there are no bulk quasiparticles.  
In the IR limit, the system decouples into two independent discs, each 
with no bulk quasiparticles. Thus the 
partition function is  
$Z_{0}=\chi_0^{}(q)$ for the UV, and 
$$Z_{0,0}=\chi_0^{}(q_1)\chi_0^{}(q_2)$$ for the IR. 
We have denoted $q_1=\exp(-2\pi\beta/R_1)$ and 
$q_2=\exp(-2\pi\beta/R_2)$, where $R_1$ and $R_2$ are the sizes 
of the discs into which the point contact splits the system in the IR
limit. The universal piece of the entropy extracted from $\chi_a(q)$  
is independent of how $q$ is taken to 1, so 
we obtain entropies of $-\ln({\cal D})$ for the entropy associated 
with each of the two disks. 
Thus from $Z_{0,0}$ we obtain  
${\cal S}_{IR}=-2\ln({\cal D})$, as required.

When there are bulk quasiparticles, they must end up on one side or 
the other once the disk is split. The UV holographic partition 
function for a quasiparticle of type $a$ is denoted $\chi_a(q)$, which 
from (\ref{Stot}) gives the entanglement entropy of ${\cal 
S}_{UV}=\ln(d_a) - \ln({\cal D})$. When there is one bulk 
quasiparticle, it must end up on one half or another, but it 
cannot end in both. There are therefore two possibilities for the IR 
partition function: 
$$Z_{a,0} = \chi_a(q_1)\chi_0^{}(q_2),\qquad\hbox{ or }\qquad  
Z_{0,a} = \chi_0^{}(q_1)\chi_a(q_2).$$ 
We cannot predict which one is realized, but one or the other will 
happen. Either of the two yields the correct entanglement entropy 
$${\cal S}_{IR}=\ln(d_a)-2\ln({\cal D}).$$  
 
\begin{figure}
\label{fig:split}
\includegraphics[width= .4\textwidth]{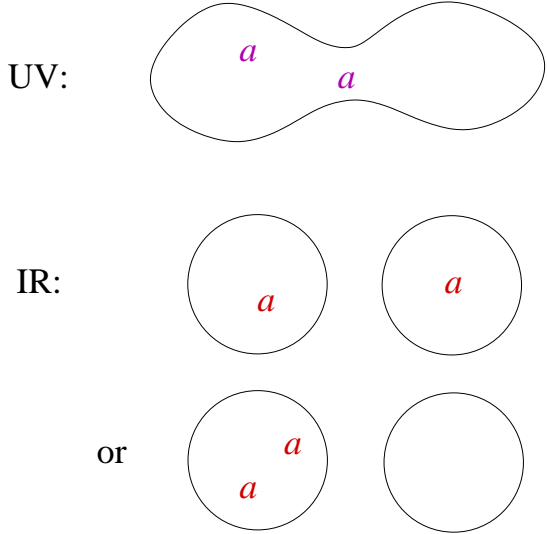} 
\caption{In the UV, bulk quasiparticles can be anywhere. In the IR,
  they must end up on one of the two sides. For two bulk
  quasiparticles of type $a$, they can both end up on one side, or on
  separate sides.}
\end{figure}
For multiple particles, we must apply the fusion rules just as we 
did for the UV holographic partition functions. 
For two quasiparticles of 
type $a$, there are three possibilities for how they might end 
up, as illustrated schematically in figure \ref{fig:split}.
If one quasiparticle ends up on each disk, then the IR holographic 
partition is $Z_{a,a} = \chi_a(q_1)\chi_{a}(q_2).$ If both 
quasiparticles end up on the same disk, then e.g.\  
$$Z_{2a,0} = Z_{2a}(q_1)Z_0(q_2)= \left(\sum_b N_{aa}^b 
\chi_b(q_1)\right) \chi_0^{}(q_2).$$  
For example, for the $p+ip$ 
superconductor with two bulk $\sigma$ quasiparticles, we have 
\begin{eqnarray*} 
Z_{2\sigma,0} &=& (\chi_I(q_1)+\chi_\psi(q_1))\chi_I(q_2),\\ 
Z_{\sigma,\sigma}&=&\chi_\sigma(q_1)\chi_\sigma(q_2). 
\end{eqnarray*} 
Thus the general rule is simple: for a given subdivision, 
the holographic partition on each half is the same as it would be for 
an isolated disc. Then multiply the characters for the two separated halves 
together.  
 
With this holographic partition for the IR, the bulk part of the 
entropy remains $\sum_a n_a \ln(d_a)$ in the IR.  It does not matter 
which half the bulk quasiparticle is on: it still contributes 
$\ln(d_a)$ in accord with the intuition that a point contact should 
not affect the bulk of the sample. As we have seen in (\ref{SUVIR}), 
the edge entropy changes, doubling from $-\ln({\cal D})$ to 
$-2\ln({\cal D})$.

\section{Discussion}

The ideas presented here grew out of our attempts to understand
quasiparticle tunneling at a point contact in a Moore-Read Pfaffian
non-Abelian quantum Hall state \cite{Fendley06a,Fendley06b}. A key
technical challenge in this problem was to formulate a bosonized
description of the tunneling of twist fields (or Ising spin fields)
$\sigma$ between Majorana fermion edge modes. We found that this could
be done if one introduced a spin-$1/2$ degree of freedom at the point
contact which served a `bookkeeping' purpose of keeping track of how
many $\sigma$ fields had tunneled.  We thereby saw that the tunneling
Hamiltonian for this problem could be mapped onto various versions
(depending on filling fraction of the quantum Hall state) of the Kondo
model.  Therefore, the issue of entropy loss due to the point contact
was inescapable, although the same issue arises in simpler cases such
as Abelian quantum Hall states but could be ignored there.  Our
interpretation\cite{Fendley06a}, on which we have elaborated here, is
that this entropy loss cannot be ascribed to the point contact {\em
per se}. Rather, it should be understood as a $(2$+$1)$-dimensional
entanglement entropy. When tunneling at a point contact causes a Hall
droplet to be effectively broken in two, the topological entanglement
entropy between the two sides becomes actual thermodynamic
entropy. Since this entropy is negative, ${\cal S}=-\ln{\cal D}$, the
entropy of the system decreases.

One implication of our results is that an attempt to formulate a tractable
representation of tunneling
in other non-Abelian quantum Hall states, such as the
Read-Rezayi states, will require introducing
an `impurity' degree of freedom at the point contact,
analogous to the Kondo spin in the Moore-Read Pfaffian
case, to give us the correct entropy loss.

It would be interesting to see which of these results apply to
gapless 2$+$1 dimensional theories in a topological phase. Such
theories are often closely related to conformal field theories, but
these theories are not describing the edge modes, but rather the
behavior in two-dimensional space. The entanglement entropy for such
systems \cite{Fradkin06} is quite similar to the entropies discussed here,
and it would be interesting to pursue the analogies further.

It would also be interesting to explore the connection of these
results with those motivated by gauge theory, string theory, and black
holes. Results quite analogous to ours have been derived using the
AdS/CFT correspondence \cite{Ryu06}. The conformal field
theories here are simpler than those which generally arise from the
AdS/CFT correspondence (ours are unitary and rational), and it would be
interesting to understand how quantum Hall physics fits into this more
general setting.

\bigskip
 
\noindent{\em Acknowledgments}\qquad 
We would like to thank J.\ Cardy, E.\ Fradkin, M.\ Freedman, A.\ Kitaev, 
A.\ Ludwig, J.\ Preskill, and N.\ Read for discussions. 
This research has been supported by the NSF under grants  
DMR-0412956 (P.F.), PHY-9907949 and DMR-0529399 (M.P.A.F.) and 
DMR-0411800 (C.N.), and by the ARO under grant W911NF-04-1-0236 (C.N.).

\appendix

\section{Examples of Quantum Dimensions} 
\label{sec:examples}

\subsection{Ising model/$p+ip$ superconductor} 
 
The Ising conformal field theory is equivalent to a free massless 
Majorana fermion. The partition function on the torus -- and hence the 
characters of interest -- can be computed using the properties of
free fermions; the main subtlety is that to compute characters for all the 
primary fields $I,\psi,\sigma$ one needs to utilize both periodic and 
antiperiodic boundary conditions on the fermion. The connection 
between the different characters and the different boundary conditions 
is reviewed in depth in, for instance, 
ref. \onlinecite{Ginsparg89}. In the $p+ip$ superconductor, inserting 
a $\sigma$ quasiparticle (i.e.\ a vortex) in the bulk flips the 
boundary conditions on the fermion between periodic and antiperiodic 
around the disk (see e.g.\ ref.\ \onlinecite{Fendley06b}). 
 
It is natural and convenient to write conformal-field-theory 
characters in terms of Jacobi theta functions. The behavior of theta 
functions under modular transformations is simple, so it is then 
straightforward to work out the modular $S$ matrix. Explicit 
expressions for the characters can be found in Refs. 
\onlinecite{Cardy86a,Ginsparg89}.  The modular $S$ matrix for the Ising 
model is \cite{Cardy86a} 
\begin{equation} 
{S}_{\rm Ising}=\begin{pmatrix} 
\frac{1}{2}&\frac{1}{2}&\frac{1}{\sqrt{2}}\\ 
\frac{1}{2}&\frac{1}{2}&-\frac{1}{\sqrt{2}}\\ 
\frac{1}{\sqrt{2}}&-\frac{1}{\sqrt{2}}&0 
\end{pmatrix} 
\label{SIsing} 
\end{equation} 
where we write the fields in the order $I,\psi,\sigma$. 
Note that $S^2=1$, as required.  
 
Using (\ref{quantdim}), we find $d_I=1$, $d_\psi=1$ and 
$d_\sigma=\sqrt{2}$, as we saw directly from the fusion rules.  Using 
(\ref{DS}), we find the total quantum dimension ${\cal D}_{p+ip}=2$, as 
we saw in (\ref{Dpp}).

\subsection{Free boson/Laughlin states} 
 
Before studying the non-abelian Moore-Read Pfaffian state, it is first useful 
to study the abelian Laughlin state at $\nu=1/3$. The latter is the 
first state in the a sequence of quantum Hall states, with the 
Moore-Read Pfaffian second. The quasiparticles and quasiholes in the 
Laughlin state at $\nu=1/3$ are abelian anyons, with charges $\pm 
1/3$. 
 
The conformal field theory describing the edge modes of the Laughlin 
states is simply a free chiral boson $\varphi(z)$ \cite{Wen91}. Here 
(and in most situations of interest), the boson is compact, meaning in 
our chiral context that operators must be invariant under the shifting 
$\varphi(z)$ to $\varphi(z) +2\pi r$ for some ``radius'' $r$ (not to 
be confused with the radius $R$ of the disk; calling $r$ a radius is a 
relic of string theory). Then standard manipulations  
show that a free boson has central charge $c=1$, and that the 
primary fields on the punctured plane under conformal symmetry 
are \cite{Ginsparg89} 
\begin{equation} 
V_\gamma(z)\equiv e^{i\gamma\varphi(z)}, \qquad\qquad \gamma=\frac{n}{r} 
\label{vertexop} 
\end{equation} 
for integer $n$. The operator $e^{i\gamma\varphi}$ appears
frequently in the Coulomb-gas approach to $2d$ statistical
mechanical models, and in string theory
as a vertex operator.
 
With the conventional normalization 
of $\varphi$, $V_\gamma$ has scaling dimension 
$\gamma^2/2$. This means that the eigenvalue of ${\cal H}$ acting on 
the state created by $V_\gamma$ is 
$2\pi(\gamma^2/2-1/24)/R$. More standard calculations show that the 
operator product expansion is 
\begin{equation} 
e^{i\gamma\varphi(z)} e^{i\delta\varphi(w)} 
=(z-w)^{\gamma\delta} e^{i\gamma\varphi(z)+i\delta\varphi(w)}. 
\label{OPE} 
\end{equation} 
To obtain correlation functions on the cylinder, it is usually most 
convenient to first work them out on the punctured plane, and then 
conformally transform them to the cylinder. In general, a primary 
field $\Phi(z,\overline z)$ of dimensions $(h,\overline h)$ transforms 
under the conformal transformation $z'=f(z)$ as (again, see e.g.\ \cite{Ginsparg89}) 
$$\Phi(z,\overline z) \to  
\left(\frac{\partial{f}}{\partial{z}}\right)^{{h}} 
\left(\frac{\partial\overline{f}}{\partial\overline{z}}\right)^{\overline{h}} 
\Phi\left(f(z),\overline{f}(\overline{z})\right).$$  
For the transformation from the punctured plane to the cylinder, we 
have $\tau+i\sigma = R\ln(z)/2\pi$, so for the operators 
$V_\gamma(z)$, we have 
\begin{equation} 
V_\gamma(z) \to \left(\frac{R}{2\pi z}\right)^{\gamma^2/2} 
V_\gamma(\tau+i\sigma). 
\label{planecylinder} 
\end{equation}

The fact that there is only one primary field on the right-hand-side 
of the operator product expansion (\ref{OPE}) means that the quantum 
dimension of all the operators $V_\gamma$ are $d_\gamma=1$.  Computing 
the total quantum dimension ${\cal D}$ for a free boson is more work, 
because there are an infinite number of primaries if we use only 
conformal symmetry since the operator product expansion for $N$ 
operators $V_{\gamma}$ gives the operator $V_{N\gamma}$. One 
interesting symmetry generator is the $U(1)$ current 
$\partial_z\varphi$.  In the fractional quantum Hall effect, this is 
the electrical charge current.  To define ${\cal D}$, we must find the largest 
possible symmetry algebra.  For any rational $r^2$, one can extend the symmetry 
algebra by some integer-dimension operator and obtain a finite number 
of primary fields, thus giving a rational conformal field theory. 
 
However, for the $\nu=1/m$ Laughlin states in the fractional quantum
Hall effect, the extended symmetry is even larger. As we discussed
above, one needs to extend the algebra by the electron
annihilation/creation operators. These have dimension $m/2$
\cite{Wen91}. A field with half-integer dimension is fermionic;
symmetries with such generators occur for example in the Gross-Neveu
model for an odd number of Majorana fermions
\cite{Witten78}. (Incidentally, in these Gross-Neveu models, the massive
quasiparticles end up having a quantum dimension of $\sqrt{2}$. \cite{Fendley02})  For the simplest non-trivial case $m=3$,
the generators are of dimension 3/2. The symmetry here is {\em
supersymmetry} \cite{Milovanovic96}.  It is called ${\cal N}=2$
supersymmetry, because there are two generators: the electron creation
and annihilation operators. The entire sequence of Read-Rezayi states
\cite{Read99}, of which the $\nu=1/3$ Laughlin state is the first
member and the Moore-Read state the second, has ${\cal N}=2$
supersymmetry as its extended symmetry.
 
Using the 
operator product expansion, it is simple to see what the primary 
fields are for the $\nu=1/3$ Laughlin state. The supersymmetry 
generators are $G^\pm=e^{\pm i\sqrt{3}\varphi}$, which indeed have 
dimension $3/2$ and charge $\pm 1$. From (\ref{OPE}), we have 
\begin{equation} 
G^\pm(z)V_\gamma(w)= (z-w)^{\pm \sqrt{3}\gamma} 
V_{\gamma \pm\sqrt{3}}(w)\left(1\ +\ \dots\right) 
\label{GOPE} 
\end{equation} 
 where the terms omitted 
are regular in $z-w$. The transformation rule (\ref{planecylinder}) 
means that when $G^\pm(z)$ is transformed from the punctured plane to 
the cylinder, it is multiplied by a factor of $z^{3/2}$. Thus the 
antiperiodicity of the electron/supersymmetry generators on the 
cylinder means that they are periodic on the punctured plane. Since we 
are only interested in this sector, the primary fields are those for 
which the exponent of the $(z-w)$ piece in (\ref{GOPE}) is an integer.  
The lowest non-zero value of 
$|\gamma|$ for which this is true is $\gamma=1/\sqrt{3}$, so we must 
have $r=1/\sqrt{3}$. Thus the allowed operators $V_\gamma$ here must 
have $\gamma=n/\sqrt{3}$ for integer $n$. In this language, the 
supersymmetry generators themselves correspond to $n=\pm 3$. Thus the 
only primary fields under the extended algebra have $n=0$ (the 
identity field), and $n=\pm 1$ (which have dimension $1/6$). The 
remaining values of $n$ can be obtained by acting with $G^\pm$ on 
these three primaries, as is clear from the operator product expansion 
(\ref{GOPE}). For example, one obtains $V_{2/\sqrt{3}}$ by acting with 
$G_+$ on $V_{-1/\sqrt{3}}$.  Physically, this means that all the edge 
modes can be obtained from these three by attaching a hole or an 
electron.  There are thus three types of quasiparticles in the 
Laughlin state at $\nu =1/3$, so 
${\cal D}_{1/3}=\sqrt{3}.$ For general integer $m=1/\nu$, the extended 
symmetry generator is $V_{\sqrt{m}}$, so by the same type of argument, 
one finds $m$ primary fields. This yields 
\begin{equation} 
{\cal D}_{\nu=1/m}=\sqrt{m} 
\label{DLaughlin} 
\end{equation} 
as noted in the introduction. 
 
One important subtlety to note is that this is the sector of the 
theory corresponding to antiperiodic boundary conditions around the 
cylinder for fermionic fields such as the supersymmetry generator. In the 
language of supersymmetry, this is called the Neveu-Schwarz sector. As 
we noted previously, periodic boundary conditions on fermions in an 
edge theory arise from a vortex in the bulk. Fields local with respect 
to a supersymmetry generator with periodic boundary conditions are in 
the Ramond sector. In an ${\cal N}=2$ supersymmetric theory, there 
must be the same number of primary fields in the Ramond sector as the 
Neveu-Schwarz sector, so ${\cal D}$ remains the same.  For the boson 
with $r=1/\sqrt{3}$, the fields $V_{n/\sqrt{3}}$ for half-integer $n$ 
are local with respect to $G_{\pm}$ with periodic boundary conditions. 
The Ramond primaries are then those with $n=\pm 1/2$ and $n=3/2$, so 
${\cal D}$ indeed remains $\sqrt{3}$. Note that $n=-3/2$ is not a primary 
because it can be obtained by acting on $V_{\sqrt{3}/2}$ with $G_-$.

Models with non-abelian statistics need not have $d_a=1$ 
like the free boson does. Thus one needs to work out the fusion 
algebra, which is simplest to do by using the modular $S$ 
matrix. Since many interesting models (e.g.\ all models of the 
fractional quantum Hall effect) involve a free boson, we display here 
the characters for the free boson. This also will provide a check on 
the above result for ${\cal D}$. 
 
For a free boson, the highest-weight states under the symmetry algebra 
including the $U(1)$ current $\partial\varphi$ are created by the 
primary fields $V_\gamma$ defined in (\ref{vertexop}). The 
highest-weight state created by $V_\gamma$ has dimension $\gamma^2/2$, 
so the leading term in the character must have exponent $\gamma^2/2 - 
1/24$.  The full characters with this symmetry algebra are (see e.g.\ Ref.
\onlinecite{Ginsparg89}) 
\begin{equation} 
\chi_\gamma(q) =  
\begin{cases} 
\frac{q^{\gamma^2/2}}{\eta} \qquad\quad&\gamma\ne p/\sqrt{2} \\ 
\frac{q^{p^2/4}-q^{(p+2)^2/4}}{\eta} \qquad&\gamma= p/\sqrt{2} \\ 
\end{cases} 
\end{equation} 
for $p$ integer. The denominator is the Dedekind eta function, which 
is defined as 
$$\eta(q)\equiv q^{1/24}\prod_{n=1}^\infty(1-q^n).$$ 
Its key property is that when $q^{-1/24}\eta$ is expanded in 
powers of $q$, the coefficient of $q^m$ is the number of partitions of 
$m$, i.e.\ the number of ways $m$ can be written as the sum of positive 
integers. This often occurs in characters, because descendant 
states are given by acting on primary states with the $L_{-n}$s for $n$ 
a positive integer.

The characters for the symmetry algebra extended to 
include supersymmetry are a sum over the characters 
$\chi_{m/\sqrt{3}}$, because the supersymmetry generators $V_{\pm 
\sqrt{3}}$ acting on $V_{m/\sqrt{3}}$ take it to $V_{(m\pm 
3)/\sqrt{3}}$, as shown in (\ref{GOPE}). To avoid confusion with the 
Ising characters and with the $\chi_\gamma$, we denote the character 
in the identity sector here as $\chi[0]$, and those for the primaries 
with $m=\pm 1$ as $\chi[\pm 1/3]$. One then finds that 
\begin{eqnarray*} 
\chi[0]&=& \sum_{n=-\infty}^\infty \chi_{n\sqrt{3}} = \frac{J(q^{3})}{\eta}\\ 
\chi[\pm 1/3] &=& \frac{1}{2}\sum_{n=-\infty}^\infty (\chi_{n/\sqrt{3}}\ 
-\chi_{n\sqrt{3}})= \frac{J(q^{1/3})-J(q^{3})}{2\eta} 
\end{eqnarray*} 
where we have defined 
$$J(p) \equiv \sum_{n=-\infty}^\infty p^{n^2/2}, 
$$ 
so that for small $p$, $J(p)=1+2p^{1/2} + 2p^2+\dots$. 
It is a straightforward exercise to generalize these free-boson 
characters to the $\nu=1/m$ case, where the extended-symmetry 
generators are of dimension $m/2$.

To find the modular $S$ matrix, we need to rewrite the characters as 
functions of $\widetilde{q}$. Luckily, doing such {\em modular 
transformations} is a well-understood branch of mathematics. We have 
defined $J(p)$ as a special case of a Jacobi elliptic theta 
function. Then, from standard mathematical 
texts, one finds that 
\begin{eqnarray} 
\nonumber 
J(\widetilde{q}^{1/\Delta}) &=& (-\Delta\ln q/2\pi)^{1/2} J(q^\Delta)\
,\\ 
\eta(\widetilde{q}) &=& (-\ln q/2\pi)^{1/2} \eta(q) \ .
\label{modtransforms} 
\end{eqnarray} 
A little algebra then gives the modular transformations 
\begin{eqnarray*} 
\chi[0](\widetilde{q}) &=& \frac{1}{\sqrt{3}\eta(q)} J(q^{1/3})\ ,  \\ 
\chi[\pm 1/3](\widetilde{q}) &=&  
\frac{\sqrt{3}}{2\eta(q)} J(q^{3}) -\frac{1}{2\sqrt{3}\eta(q)}
J(q^{1/3})\ .  
\end{eqnarray*} 
We now need to rewrite the right-hand-side 
in terms of $\chi[0](q)$ and $\chi[\pm 1/3](q)$. We can exploit the 
equality $\chi[1/3](q)=\chi[-1/3](q)$ to find the modular $S$ matrix which is both 
symmetric and unitary. We find 
$${S}_{\nu=1/3} = 
\begin{pmatrix} 
\frac{1}{\sqrt{3}}& \frac{1}{\sqrt{3}}&  
\frac{1}{\sqrt{3}}\\ 
\frac{1}{\sqrt{3}}& \frac{1}{2}-\frac{1}{2\sqrt{3}}& \ 
-\frac{1}{2}-\frac{1}{2\sqrt{3}}\\ 
\frac{1}{\sqrt{3}}& -\frac{1}{2}-\frac{1}{2\sqrt{3}} 
&\frac{1}{2}-\frac{1}{2\sqrt{3}}
\end{pmatrix} 
$$ 
{}From this modular $S$ matrix we can read off the quantum dimensions 
$d_a=1$ for all three quasiparticles, and ${\cal D}=1/S_{0}^0 =\sqrt{3}$, 
in agreement with our earlier results. 
 
\subsection{Moore-Read state} 
 
The edge theory for the $\nu=5/2$ Moore-Read state consists of a free 
boson (usually called the charge mode), and a Majorana fermion 
(usually called the neutral mode) \cite{Milovanovic96}. This conformal 
field theory has $c=3/2$. 
The edge electron/hole creation 
operators (i.e.\ the supersymmetry generators) are 
$$G^\pm = \psi\, e^{\pm i \sqrt{2} \varphi}.$$ Because the fermion 
$\psi$ has dimension $1/2$, the generators have dimension $3/2$, and 
it is simple to show that as in the case of the $\nu=1/3$ Laughlin state, the 
extended symmetry is ${\cal N}=2$ supersymmetry. 
 
The primaries of the combined field theory 
must be products of the primaries $I,\psi,\sigma$ of the Ising model 
with the primaries $V_\gamma$ for a free boson. The Ising primary 
fields have dimensions $0,1/2,1/16$ respectively, so by using the OPE 
(\ref{OPE}) and the Ising fusion rules (\ref{Isingfusion}), it is 
simple to work out the powers of $z-w$ in the operator products in the 
combined theory. Demanding that the electron/superconformal 
generators be periodic on the punctured plane gives the primaries of 
the extended symmetry algebra to be 
\begin{equation} 
I,\ \sigma V_{\pm 1/(2\sqrt{2})},\ \psi,\ V_{\pm 1/\sqrt{2}}. 
\label{MRprimaries} 
\end{equation} 
All primaries are all invariant under the combined transformations 
$\sigma\to -\sigma$, $\psi\to \psi$ and $\varphi\to \varphi+ 
2\pi\sqrt{2}$; this is the condition that ensures locality. Fields 
that are odd under this symmetry comprise the Ramond sector.

We can compute the quantum dimensions without the full computation of 
the modular $S$ matrix. The reason is that in (\ref{MRprimaries}) we 
have decomposed the operators into products of Ising fields with 
bosonic operators, and we know the fusion algebra of both.  
Since the boson and the Ising theory are independent, the quantum 
dimension of the product of fields from the two is just the product of 
the two quantum dimensions. Namely, 
$I$, $\psi$ and $V_\gamma$ have $d_I=d_\psi=d_\gamma=1$, while 
$d_\sigma=\sqrt{2}$. The total quantum dimension for the Moore-Read 
state is thus 
\begin{equation} 
{\cal D}_{MR} = \sqrt{1+2+2+1+1+1} =2\sqrt{2}. 
\label{DMR} 
\end{equation} 
 
To check this, we find the characters for the fields in the 
Moore-Read state. These involve both the Ising characters and the 
free-boson characters, but are not just simple products. The reason is that 
we want only the states  
in the antiperiodic (Neveu-Schwarz) sector, which are invariant under 
$\varphi\to\varphi+2\pi\sqrt{2}$, $\sigma\to -\sigma$. We want to 
include descendants which are found by acting with the supersymmetry 
generators $G^\pm$, which is invariant under the symmetry. The characters 
for the six primary fields listed in (\ref{MRprimaries}) are then 
\begin{eqnarray*} 
I:&\qquad& \chi_I\, \frac{J(q^8)}{\eta} +  
\chi_{\psi}\frac{J(q^2)-J(q^8)}{\eta}\\ 
\sigma V_{\pm 1/2\sqrt{2}}: && \chi_\sigma\frac{J(q^{1/8}) - 
J(q^{1/2})}{2\eta}\\ 
\psi:&& \chi_\psi\, \frac{J(q^8)}{\eta} + \chi_{I}\frac{J(q^2)-J(q^8)}{\eta}\\ 
V_{\pm 1/\sqrt{2}}:&& (\chi_I +\chi_{\psi})\frac{J(q^{1/2})-J(q^{2})}{2\eta}  
\end{eqnarray*} 
Using the Ising modular $S$ matrix and the modular transformations in 
(\ref{modtransforms}) one can work out the Moore-Read modular $S$ 
matrix. For the quantum dimensions, we only need the first row, which 
is 
$$S^a_0 = \left(\frac{1}{2\sqrt{2}},\frac{1}{2},\frac{1}{2}, 
\frac{1}{2\sqrt{2}},\frac{1}{2\sqrt{2}},\frac{1}{2\sqrt{2}}\right).$$ 
We recover of course the individual quantum dimensions of $\sqrt{2}$ 
for the two quasiparticles involving the $\sigma$ field, and $1$ for 
the other four, as well as the total quantum dimension ${\cal 
D}_{MR}=1/S_0^0=2\sqrt{2}$.

Just to test the formalism a little more, let us compute the quantum 
dimensions for the Moore-Read state for a system of {\em bosons} at $\nu=1$.
The edge theory is very similar, except now the 
extended symmetry is not supersymmetry, but rather $SU(2)$.
The conformal field theory is still an Ising model and a free boson,
except now the boson is at a different radius. At this radius,
the boson can be fermionized into a free Dirac fermion. This can be
split into two free Majorana fermions which, combined with the
Majorana fermion from the neutral sector, form an $SU(2)$
triplet \cite{Fradkin98}. This model is equivalent, via
non-Abelian bosonization, to the $SU(2)_2$ Wess-Zumino-Witten 
model, with the extended 
symmetry being the corresponding Kac-Moody algebra.
The extra $SU(2)$ generators are 
$$ J_\pm = \psi\, e^{\pm i\varphi}$$ 
while $J_z$ is the $U(1)$ current $\partial\phi$. Since the theory is 
bosonic, the symmetry generators are periodic on the cylinder; because 
$J_\pm$ and $J_z$ have dimension $1$, they are periodic on the 
punctured plane as well. The primary fields are therefore 
$$I,\ \sigma V_{1/2},\ \psi$$  
The total 
quantum dimension of this theory is therefore 
$${\cal D}_{SU(2)_2} = \sqrt{1+2+1} =2.$$ 
Note that $\sigma V_{-1/2}$ is not primary because it 
can be obtained by acting with $J_-$ on $\sigma V_{1/2}$. This is just 
a fancy way of saying that the two form a doublet under 
$SU(2)$. Likewise, the fields $\psi,\ e^{i\varphi},\ e^{-i\varphi}$ 
form a triplet under $SU(2)$.  
 
\subsection{Read-Rezayi} 
 
For a level-$k$th bosonic Read-Rezayi state \cite{Read99},
the edge conformal field theory is 
the Wess-Zumino-Witten model $SU(2)_k$. This has been well studied in 
the literature, and the modular $S$ matrix is given by 
\cite{Gepner86} 
$$ S_a^b=\sqrt{\frac{2}{k+2}} \sin\left(\frac{\pi (a+1) (b+1)}{k+2}\right)$$ 
for $0\le  a,b\le k$. The quantum dimensions are therefore 
\begin{equation} 
d_a=\frac{\sin(\pi (a+1)/(k+2))}{\sin(\pi/(k+2))} 
\label{su2dim} 
\end{equation} 
and the total quantum dimension is 
\begin{equation} 
{\cal D}_{SU(2)_k} = \frac{\sqrt{k+2}}{\sqrt{2}\sin(\pi/(k+2))}. 
\label{su2totaldim} 
\end{equation} 
 
For the Read-Rezayi fermionic fractional quantum Hall states, we
cannot instantly read off the modular $S$ matrix from the
literature. The reason is that we need the characters for the extended
symmetry in the Neveu-Schwarz sector of the ${\cal N}=2$
supersymmetric minimal models, which do not seem to be in the
literature. However, we can compute the quantum dimensions by
utilizing the fact that the ${\cal N}=2$ primaries are closely related
to $SU(2)_k$ fields. Namely, the $SU(2)_k$ fields are labeled by their
$SU(2)$ quantum numbers, a spin $l/2$ with $l=0,1,\dots k$ and a
charge $m=-l,-l+2, \dots, l$. The primary fields in the Neveu-Schwarz
sector of the superconformal field theory are labeled in the same way,
and differ only in multiplying by a field involving the charge boson,
namely $V_{\gamma}$ with
$\gamma=-m/\sqrt{2(k+2)}$. \cite{Boucher86,Zamo86} This changes the
field's dimension, but not the fusion rules, because $m$ is a
conserved quantum number in fusion (in the Hall effect the physical
charge of the quasiparticle created by the field is $me/(k+2)$). Thus
a given field's quantum dimension depends only on the index $l$, and
all the quantum dimensions of the Read-Rezayi fractional quantum Hall
states are given by (\ref{su2dim}).  There are more primary fields in
the supersymmetric case, because in the $SU(2)_k$ case, the $l+1$
fields with a given $l$ form an $SU(2)$ multiplet, and so correspond
to only one primary field of the $SU(2)$-extended algebra. The total
quantum dimension for the $k$th Read-Rezayi state for the fractional
quantum Hall effect is therefore
\begin{eqnarray}
\nonumber 
{\cal D}_{\hbox{RR}} &=&\left(\sum_{l=0}^{k} (l+1)  
\frac{\sin^2(\pi (l+1)/(k+2))}{\sin^2(\pi/(k+2))}\right)^{1/2}\\
&=& \frac{k+2}{2\sin(\pi/(k+2))}\ .
\label{DRR} 
\end{eqnarray}

\vskip -0.5cm 
 
 

\end{document}